# Plasma Adiabatic Lapse Rate

Peter Amendt, Claudio Bellei and Scott Wilks

*Lawrence Livermore National Laboratory, Livermore, CA 94551 USA*

The plasma analog of an adiabatic lapse rate (or temperature variation with height) in atmospheric physics is obtained. A new source of plasma temperature gradient in a binary ion species mixture is found that is proportional to the concentration gradient $\bar{\nabla}\alpha$ and difference in average ionization states $Z_2 - Z_1$. Application to inertial-confinement-fusion implosions indicates a potentially strong effect in plastic (CH) ablators that is not modeled with mainline (single-fluid) simulations. An associated plasma thermodiffusion coefficient is derived, and charge-state diffusion in a single-species plasma is also predicted.

The adiabatic lapse rate is a well-understood phenomenon in atmospheric physics that leads to a predicted temperature decrease of 9.8 °C per kilometer of height in a dry, accelerated atmosphere in hydrostatic equilibrium.[1] The associated temperature gradient is expected to lead to diffusive molecular species separation at the upper levels (>120 km) of the atmosphere where molecular diffusion dominates over turbulent diffusion.

The notion of an adiabatic lapse rate has not been applied to the plasma state, much less inertial-confinement-fusion (ICF) implosions of spherical capsules filled with thermonuclear fuels, e.g., deuterium (D) and tritium (T). An imploding ICF shell also resembles an accelerating atmosphere (ahead of the ablation front) so that an adiabatic lapse rate notionally applies. An intensive effort is underway at the National Ignition Facility (NIF) to demonstrate ignition of DT fuels in the indirect-drive configuration where laser light is converted to soft x rays in a high-$Z$ cylindrical enclosure (or hohlraum) that bathe a centrally-located plastic (CH) capsule.[2] Internal pressure gradients self-consistently arise from the acceleration of the imploding shell in direct reaction to the x-ray ablation, but the associated temperature gradients are often neglected in analytical treatments – in contrast to their inclusion in radiation-hydrodynamics (RH) simulations.[3] Temperature gradients provide an additional source for mass diffusion, i.e., thermodiffusion, beyond pressure gradients ("barodiffusion") and concentration gradients (classical diffusion). Binary mass diffusion may be important for ICF as the spatial segregation of the fuel constituents may lead to anomalous behavior and degraded performance.[4] The strength of thermodiffusion has not been assessed for ICF plasmas, largely because the temperature gradient has heretofore not been analytically assessed. The derivation of a plasma adiabatic lapse rate in this Letter identifies a new plasma-based source of temperature gradient (and resulting heat flow) proportional to the product of the concentration gradient and difference in average ionization states between the two ion species. A significant departure from single-fluid, (RH) simulated temperature profiles is found for mixed-species ablators. For a plastic (CH) ablator, the associated temperature gradient modification during shell acceleration $\vec{g}$ may overshadow the believed uncertainties in thermal conductivity models used in ICF research. The immediate impact of significantly reduced temperature gradients in the CH ablator compared with mainline RH simulations is a potential for increased hydrodynamic instability on the fuel-ablator interface and reduced thermal conduction in the x-ray preheated, buried, Ge-doped CH layer, leading to greater volume expansion. Knowledge of the temperature gradient is then used to derive a thermodiffusion coefficient in an ideal plasma mixture. A novel and important source of binary mass diffusion (or "electrodiffusion") driven by an electric field is found, varying as the difference in ion charge-to-mass ratios. A new phenomenon of charge-state diffusion within a single-species plasma is also predicted, whereby the higher (lower) charge states diffuse in a direction opposite to (along) the shell acceleration. These sources of diffusion may collectively lead to a significant species separation or redistribution, especially near the center of the fuel (following shock rebound from the origin) where the ion mean-free-paths are relatively large. Species separation may also have important astrophysical implications for the dynamics of the expanding (initially high Mach number ~1000) shock wave ahead of the supernova ejecta as the interstellar medium is swept up.

Consider a steady-state plasma of two ion species with each component labeled by the subscripts $i$=1,2. Assuming ideal equations-of-state for each ion species, the total pressure, including the free electron contribution, is given by $P = nk_B T(1+\bar{Z}T_e/T)$, where $n = n_1 + n_2$ is the total ion number density, $k_B$ is Boltzmann's constant, $T$ is the (equal) temperature for each ion species, $T_e$ is the electron temperature, $\bar{Z} = Z_2 - f\Delta Z$ is the average ionization state of the mixture, $f \equiv n_1/n$ is the fractional number density of species 1, and $\Delta Z = Z_2 - Z_1 \geq 0$. In a plasma, the energy exchange between ion species proceeds at a much faster rate than electron-ion coupling, so the allowance of unequal ion and electron temperatures is justified. Upon forming the differential of $P$ and assuming an adiabatic equation-of-state for the composite plasma mixture, we have $(\gamma-1)/\gamma \cdot d\ln P = d\ln T - df \cdot (T_e/T)\Delta Z/(1+\bar{Z}T_e/T)$, where $\gamma$ is the ratio of specific heats. Introducing the light-ion mass fraction $\alpha \equiv m_1 n_1/(m_1 n_1 + m_2 n_2)$, gives for the plasma adiabatic lapse rate:

$$\nabla \ln T = \nabla \ln P \cdot \left(\frac{\gamma-1}{\gamma}\right) + \frac{\Delta Z}{\bar{Z}+T/T_e} \frac{m_1}{m_2} \frac{\nabla \alpha}{\left[\alpha + (1-\alpha)m_1/m_2\right]^2}, \qquad (1)$$

where $m_i$ is the mass of the $i$-th ion species, $\nabla \ln P = -A_w \vec{g}/(1+\bar{Z}T_e/T)k_B T$ and $A_w = fm_1 + (1-f)m_2$ is the number-weighted ion mass. The last term in Eq. (1) distinguishes the plasma adiabatic lapse rate from the gas-dynamical term $[\nabla \ln P \cdot (\gamma-1)/\gamma]$, and provides a novel temperature-gradient contribution proportional to $\Delta Z \cdot \nabla \alpha$. This source term is tantamount to a gradient in the average ionization state $\bar{Z}$, which is known to generate a charge separation in an accelerating plasma.[5] The total plasma electric field follows from electron momentum balance $[E = -\nabla P_e/en_e]$,[6] and using $\nabla n/n = \nabla \rho/\rho + \nabla \alpha(m_2/m_1 - 1)/[1+\alpha(m_2/m_1 - 1)]$, where $\rho$ is the total mass density:

$$E = \left(\frac{k_B T}{e}\right)\left[-\nabla \ln P + \frac{\Delta Z \cdot \nabla \alpha}{[\alpha Z_1 m_2/m_1 + (1-\alpha)Z_2][\alpha(Z_1+1) + (1-\alpha)(Z_2+1)m_1/m_2]}\right].$$

(2)

Using momentum conservation for ion species 1 and Eq. (2) to eliminate $E$, we obtain the plasma barometric formula $\nabla \alpha = k_\alpha \nabla \ln P$ where:

$$k_\alpha = -\alpha(1-\alpha)\left[(1+Z_1 T_e/T)/m_1 - (1+Z_2 T_e/T)/m_2\right]m_1 \times$$
$$\frac{[\alpha Z_1 m_2/m_1 + (1-\alpha)Z_2][\alpha(1+Z_1 T_e/T)m_2/m_1 + (1-\alpha)(1+Z_2 T_e/T)]}{[\alpha Z_1(1+Z_1 T_e/T)m_2/m_1 + (1-\alpha)Z_2(1+Z_2 T_e/T)]}.$$

(3)

Equation (3) states that a concentration gradient (and ion species separation) must arise when an applied pressure gradient occurs over a time scale much greater than an ion-ion collision time, i.e., steady-state conditions. An exception is $D^3He$ fuels (in contrast to mainline DD and DT fuels) that have a vanishing concentration gradient for all mass fractions. Equation (3) is also independent of the adiabatic index $\gamma$, implying more general applicability under non-adiabatic conditions as in an ablation front.

Figure 1 displays the steady-state temperature- and concentration-gradient profiles from Eqs. (1-3) versus admixtures of carbon and hydrogen. The strong effect of species separation, i.e., $\nabla \alpha \neq 0$, and nonzero $\Delta Z$ on the temperature profile is apparent. For vanishing $\Delta Z$ or $\nabla \alpha$, the temperature gradient profile remains constant at 0.4 (in units of $\nabla \ln P \propto -g$) for a chosen ratio of specific heats $\gamma=5/3$ – just as mainline RH simulations would predict. However, at fairly small values of $\alpha$ and $1-\alpha$ the temperature gradient varies rapidly, changing sign and remaining negative over a wide range of $\alpha$. This case points out the importance of coupled ionization and acceleration-induced concentration gradients on altering the average hydrodynamic profiles in an ICF plasma by greatly reducing and possibly reversing the direction of conductive heat flow for a mixed species ablator - such as CH.

The atomic mixing of fuel and shell material resulting from hydrodynamic instability, e.g., Rayleigh-Taylor and Richtmyer-Meshkov, is a general feature of ICF implosions that raises ignition thresholds and reduces target performance by contaminating a significant fraction of the fuel mass with penetrating higher-$Z$ shell material.[7] The current baseline ICF target for demonstrating ignition on the NIF is a CH ablator with a buried, graded (germanium or silicon) dopant layer for x-ray preheat shielding of the encapsulated, cryogenic, main DT fuel layer.[8] Figure 2 shows key profiles for this target as a function of light ion mass fraction, physically spanning the entire "mix" layer spatial extent from pure Ge to CH. The temperature gradient is highly affected by $\Delta Z$ and $\nabla \alpha$ as for CH [Fig. 1], while the concentration gradient remains strong. In applying Eqs. (1-3), the average $Z$ of the Ge (CH) constituents is taken to be 6.0 (1.5) following multi-shock traversal, and the average ion mass is 72.3 (6.5) atomic mass units ($A$). The vertical dashed line denotes the experimental value of the maximum concentration (1.0 at.%) of buried germanium near the inside surface of the ablator. It must be noted that for both undoped and Ge-doped CH in the NIF ablator design, the operational values of $\alpha$ coincide with a nearly vanishing (or even a reversed) temperature gradient relative to baseline RH simulations. A reduced temperature gradient leads to lower conductive energy loss in the Ge-doped CH layers, resulting in greater volume expansion. In fact, recent (Zn) backlit implosion data suggest an anomalously large $\approx 2\times$ Si-doped CH layer expansion compared with RH predictions.[9,10]

Knowledge of the steady-state temperature gradient now enables an assessment of diffusive mass flow including thermodiffusion, which leads to an estimate of the time scale for establishing a steady state atmosphere. To evaluate the diffusive mass flow of the light ions $i_1$ resulting from generalized forces, we write:[11] $i_1 = -A_\mu \nabla \mu - A_T \nabla T$, where $A_\mu, A_T$ are coefficients, $\mu = \mu_1 / m_1 + \mu_2 / m_2 + (Z_1 / m_1 - Z_2 / m_2) e \Phi$ is the electrochemical potential of the mixture, $\mu_i = k_B T \ln f_i + \mu_{i0}(T)$ is the chemical potential of each species, and $\Phi$ is the electrostatic potential. After changing to the usual variables $P, T, \alpha, \Phi$, we obtain:

$$i_1 = -\rho D \left[ \nabla \alpha + k_P \nabla \ln P + k_E \frac{e \nabla \Phi}{k_B T} + k_T \nabla \ln T \right], \quad (4)$$

where $D$ is the classical diffusion coefficient,

$$k_P = P(\partial V/\partial \alpha)_{P,T,\Phi}/(\partial \mu/\partial \alpha)_{P,T,\Phi} = \alpha(1-\alpha)\left[(m_2-m_1)\left(\frac{\alpha}{m_1}+\frac{1-\alpha}{m_2}\right)(1+\bar{Z})+\Delta Z\right]$$

(5)

is the barodiffusion coefficient,[12] $V=1/\rho$ is the specific volume,

$$k_E = k_B T(Z_1/m_1 - Z_2/m_2)/(\partial \mu/\partial \alpha)_{P,T,\Phi} = \alpha(1-\alpha)\left(\frac{Z_1}{m_1}-\frac{Z_2}{m_2}\right)\left(\frac{\alpha}{m_1}+\frac{1-\alpha}{m_2}\right)m_1 m_2$$

(6)

is the "electrodiffusion" coefficient that describes an intrinsic plasma field contribution to the diffusive mass flux, and $k_T$ is the thermodiffusion coefficient. Both the barodiffusion and electrodiffusion coefficients are local thermodynamic equilibrium (LTE) quantities, reducing to the proper fluid limit[11] and vanishing, respectively, for equal charge-to-mass ratios. $k_T$ is an inherent non-LTE quantity and is determined as follows. Using Eqs. (1-3) to rewrite each gradient term in Eq. (4) in terms of $\nabla \ln P \neq 0$, setting $i_1 = 0$ in steady state, we can then solve directly for $k_T$ in terms of $k_\alpha, k_P, k_E$ [Eqs. (3,5,6)]:

$$k_T = -\frac{k_P + k_\alpha + k_E\left\{1-\dfrac{k_\alpha \Delta Z}{[\alpha Z_1 m_1/m_2 + (1-\alpha)Z_2][\alpha(1+Z_1)+(1-\alpha)(1+Z_2)m_1/m_2]}\right\}}{\dfrac{\Delta Z}{\bar{Z}+1}\dfrac{m_1}{m_2}\dfrac{k_\alpha}{[\alpha+(1-\alpha)m_1/m_2]^2}+\dfrac{\gamma-1}{\gamma}}.$$

(7)

Note that $k_T$ undergoes a resonance when the denominator ($\propto \nabla \ln T$) vanishes, cf., Eq. (1), but the mass flow contribution $(\propto k_T \nabla \ln T)$ remains bounded. However, this is not the case for the energy flux counterpart to Eq. (4):[11] $q = \left[k_T(\partial \mu/\partial \alpha)_{P,T,\Phi} - T(\partial \mu/\partial T)_{P,\alpha,\Phi} + \mu\right]i_1 - \kappa \nabla T$, where $\kappa$ is the thermal conductivity. Note that the resonant nature of $k_T$ is an intrinsic plasma effect, depending on nonzero $\Delta Z$. This feature may lead to anomalous (convective) energy transport in the Ge-doped CH layer of a NIF ablator because $\nabla \ln T \cong 0$ nearly coincides with the experimental concentration of 1.0 at.% Ge dopant, cf., Fig. 2.

Figure 3a-b shows the various diffusion coefficients for DT and D$^3$He fuels. The barodiffusion coefficient dominates over the electrodiffusion and classical diffusion coefficients, but is generally smaller (in magnitude) than the thermodiffusion coefficient. However, the pressure gradients are often significantly larger than the temperature gradients, rendering barodiffusion more important in practice. For example, a Mach number 2 ($\infty$) gas-dynamical shock has more than $2\times$ ($4\times$) stronger pressure gradients for $\gamma = 5/3$.

The condition for applying Eqs. (1-3) in ICF studies is that there is sufficient time to establish steady state conditions over an implosion time scale. We now test this assertion with use of Eq. (4) that naturally leads to an ion drift speed $v_d \equiv 2i_1/\rho \cong 2Dk_E eE/k_B T$, where $D = (\bar{Z}+1)k_B T/A_w \nu_{12}$ is the classical diffusion coefficient, $\nu_{12}$ is an ion-ion collision frequency, and the electrodiffusion term is singled out for simplicity and importance. Based on experiment[15] and supporting theory[5], $E$ is on the order of $10^4$ statVolts/cm or larger, and $v_d \approx k_E \cdot 10^{18} [cm/s^2]/\nu_{12}$ for $\bar{Z} \cong 1$ and $A_w \cong 6.5$. For $T$>20 eV and $\rho$<1 g/cm$^3$, $\nu_{12} \cong O(10^{13} s^{-1})$ or less,[16] and $v_d$ >1 μm/ns, which defines a threshold value for initiating appreciable species separation in an ICF implosion. This condition may be met near the time of shock rebound from the fuel center, in the shell material ahead of the ablation front following multiple shock passages, and behind the ablation front.

A new feature of Eq. (3) is the possibility of binary charge-state diffusion for a single ion species. Setting $m_1 = m_2$ now gives $k_\alpha > 0$:

$$k_\alpha = \alpha(1-\alpha)\Delta Z \frac{[\alpha Z_1 + (1-\alpha)Z_2][\alpha(Z_1+1)+(1-\alpha)(Z_2+1)]}{[\alpha Z_1(Z_1+1)+(1-\alpha)Z_2(Z_2+1)]} \quad (8)$$

where $\alpha$ now refers to the *number* fraction of charge state $Z_1$, and $k_E = k_P = \alpha(1-\alpha)\Delta Z$. Figure 4 shows the various diffusion coefficients for the case of pure carbon with $\Delta Z$=3,5. We note that all four sources for charge-state diffusion in Eq. 4 have the same sign and act in tandem. For the case of an inward propagating shock through pure C, the ions with charge state $Z_1$ ($Z_2$) will diffuse inward (outward). More generally, a Saha-like distribution of charge states (near LTE) will be modified accordingly in an accelerating atmosphere if the recombination and ionization rates are sufficiently small. The physical result of this effect is a potential abundance of higher-$Z$ ion states in the vicinity of the outside of the dopant layers. A higher residual $\bar{Z}$ may lead to a reduced (areal) mass ablation rate $\dot{m} \propto A/(\bar{Z}+1)$, as well as an increased albedo to drive x rays at late time.

Binary charge-state diffusion in a single-species plasma necessarily leads to larger temperature gradients. Using Eq. (8) in Eq. (1) with $m_1 = m_2$ gives:

$$\nabla \ln T = \nabla \ln P \left[ \frac{\gamma-1}{\gamma} + \frac{(\Delta Z)^2 \bar{Z}}{4\bar{Z}(1+\bar{Z})+(\Delta Z)^2} \right], \qquad (9)$$

where the binary charge states $Z_1, Z_2$ have been replaced by the principal ionization distribution moments $\bar{Z}, \Delta Z$, and $\alpha \simeq 1/2$ is used accordingly. For Saha-like equilibrium distributions, $\Delta Z$ (or FWHM of the ionization distribution) is on the order of 2-3, and the last term in Eq. (9) can be comparable to the gas-dynamical, temperature gradient source term.

The effect of a temperature gradient reversal in an ICF capsule as argued in this Letter may play a prominent role in the evolution of Rayleigh-Taylor hydrodynamic instability growth. Near the DT-CH interface well before deceleration onset, the temperature profile in the innermost CH layer adjacent to the fuel may be significantly flatter due to the influence of the $\Delta Z \cdot \nabla \alpha$ term [in Eq. 1], compared with single-fluid RH predictions[13]. However, the temperature profile must be continuous across the DT-CH interface, forcing a stronger temperature gradient in the vicinity of the interface. Conductive heat flow will then dominate over this boundary layer, giving rise to locally anti-aligned density and temperature profiles as in an ablation front where $d \ln P / d \ln T \ll 1$. The magnitude of the increase in temperature gradient is estimated as: $\nabla T_{new} = \nabla T_{old}(1 + \Delta_{CH} \eta / 2\Delta_{fuel})$, where $\Delta_{CH}$ is the width of the innermost CH layer (~10 μm), $\Delta_{DT}$ is the width of the conductively heated DT boundary layer (~ 2-3 μm), and $\eta \equiv \Delta Z \cdot k_\alpha (m_1/m_2)(\gamma-1)/\gamma(1+\bar{Z})[\alpha+(1-\alpha)m_1/m_2]^2$ is a measure of the strength of the temperature gradient across the adjacent CH layer. For a flat or decreasing temperature profile ($\eta \geq 1$) the temperature gradient enhancement can be several-fold, leading to a similar increase in the magnitude of the density gradient. An increase in the density gradient leads to less density-gradient stabilization of high modes,[7] and the potential for more growth and mix. This scenario is consistent with former studies on the effects of an uncertainty in the thermal conductivity $\kappa$: A $0.30\times$ multiplier leads to more instability growth from reduced thermal transport.[14] A main message of this Letter is that an uncertainty in the temperature profiles relative to RH predictions may be as important – if not more so – as the uncertainty in $\kappa$.

In summary, the notion of an adiabatic lapse rate is applied to a plasma and shown to lead to a novel source of temperature-gradient that depends on the concentration gradient and difference in ion charge states in a binary mixture. Application of the model shows significant plasma modifications to the temperature gradient compared with gas dynamics (upon which single-fluid RH simulations are based). The derivation of a temperature gradient is used to derive a coefficient of plasma thermodiffusion. A new type of single-species diffusion leading to a spatial redistribution of ion charge states is also shown. Finally, the implications of the modified temperature gradient for a binary species ablator on potentially enhanced hydrodynamic instability growth are described.

Useful discussions with Grigory Kagan are appreciated. This work was performed under the auspices of Lawrence Livermore National Security, LLC (LLNS) under Contract DE-AC52-07NA27344 and supported by LDRD-11-ERD-075.

FIGURE CAPTIONS

**Fig. 1:** Temperature gradient (green) and concentration gradient (blue) profiles normalized to logarithmic derivative of total pressure for $\gamma=5/3$ and $T_e=T$ plasma versus light ion mass fraction and ionization state difference $\Delta Z$.

**Fig. 2:** Temperature gradient (green), concentration gradient (blue), and electric field (red) profiles normalized to the logarithmic derivative of total pressure for a $\gamma=5/3$ plasma versus light ion mass fraction for CH-Ge mixtures (solid lines). Vertical dashed line denotes Ge-doped CH experimental concentration and dashed blue line the gas-dynamical version of $d\alpha/dx$, i.e., with $Z_1 = Z_2 = 0$.

**Fig. 3a-b**: Mass diffusion coefficients, $k_\alpha$ [Eq. 3; red], $k_P$ [Eq. 5; green], $k_E$ [Eq. 6; blue], $k_T$ [Eq. 7; black] for DT (a) and D$^3$He (b) fuel mixture versus light ion mass fraction $\alpha$.

**Fig. 4:** Ionization-state diffusion coefficients, $k_\alpha$ [Eq. 3; red], $k_P$ [Eq. 5; green], $k_T$ [Eq. 7; black] for pure carbon versus $Z_1$ charge-state concentration $\alpha$ with $Z_2=6$, $Z_1=3$.

REFERENCES

[1] J.T. Houghton, *The Physics of Atmospheres* (Cambridge University Press, Cambridge, 1986).

[2] J.D. Lindl and E.I. Moses, Phys. Plasmas **18**, 050901 (2011).

[3] M.M. Marinak *et al.*, Phys. Plasmas **8**, 2275 (2001).

[4] P. Amendt *et al.*, Phys. Rev. Lett. **105**, 115005 (2010); P. Amendt *et al.*, Phys. Plasmas **18**, 056308 (2011).

[5] P. Amendt *et al.*, Plasma Physics Controlled Fusion **51**, 124048 (2009).

[6] We ignore ignoring the contribution of the thermal force $\left[-0.71 k_B \nabla T / e \text{ for } Z=1\right]$; see S.I. Braginskii, *Reviews of Plasmas Physics*. Vol. 1, edited by M.A. Leontovich (Consultants Bureau, New York, 1965).

[7] J.D. Lindl *et al.*, Phys. Plasmas **11**, 339 (2004).

[8] S.W. Haan *et al.*, Fusion Sci. Technol. **59**, 1 (2011).

[9] Damien Hicks *et al.*, Phys. Plasmas (submitted).

[10] Evaluation of Eq. 1 for the temperature-gradient modification in a Si-doped CH mixture (2.0 at.%) shows a $\approx 2\times$ reduction.

[11] L.D. Landau and E.M. Lifshitz, *Fluid Mechanics* (Pergamon, Oxford, 1959), p. 225.

[12] The barodiffusion coefficient $k_P$ must be independent of the electric field for thermodynamic consistency, in contrast to previous work[4]. The electric field dependence in the mass diffusion flux is separate and wholly resides in the electrodiffusion term in Eq. 4.

[13] S.W. Haan *et al.*, Phys. Plasmas **18**, 051001 (2011).

[14] B.A. Hammel *et al.*, High Energy Density Phys. **6**, 171 (2010).

[15] C.K. Li *et al.*, Phys. Rev. Lett. **100**, 225001 (2008).

[16] J.D. Huba, *NRL Plasma Formulary* (1998), p. 32.

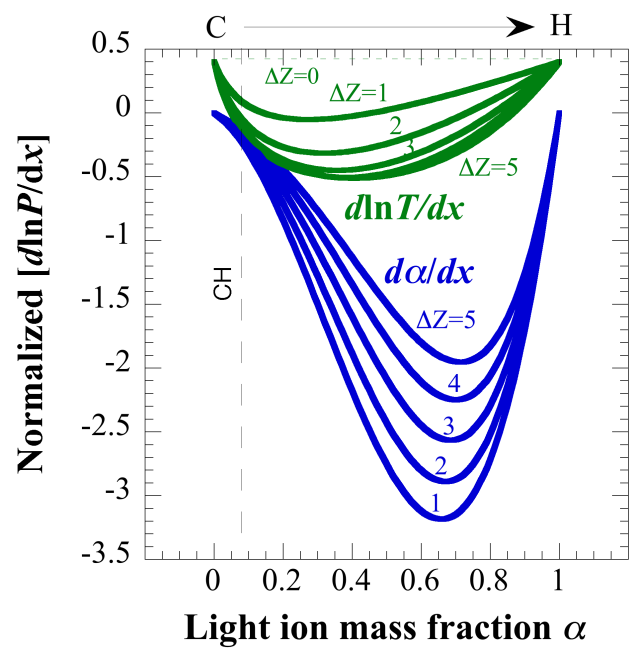

**Fig. 1**

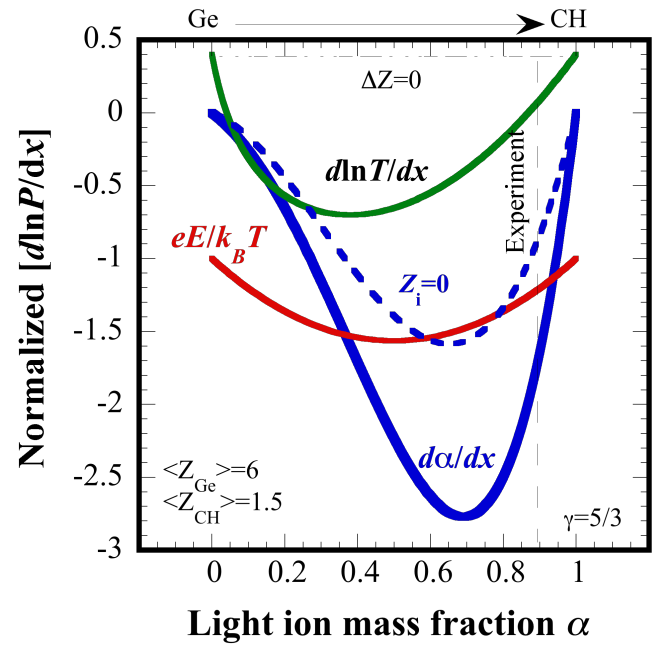

Fig. 2

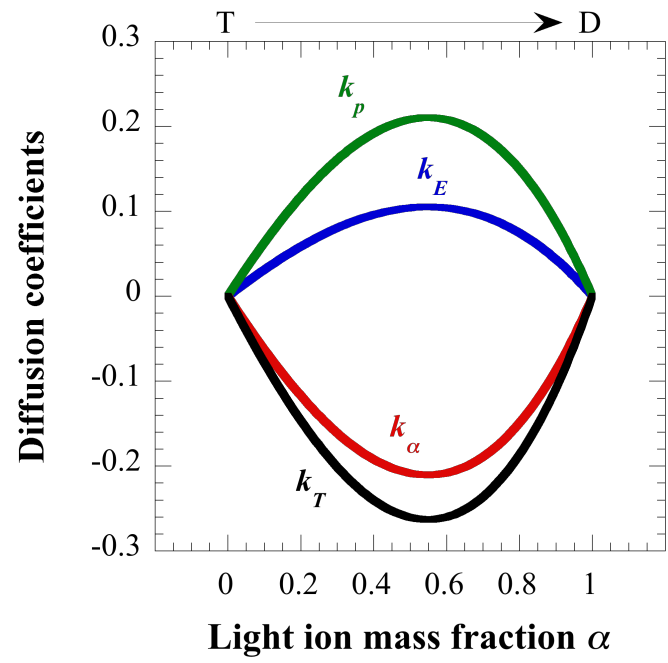

Fig. 3a

**Fig. 3b**

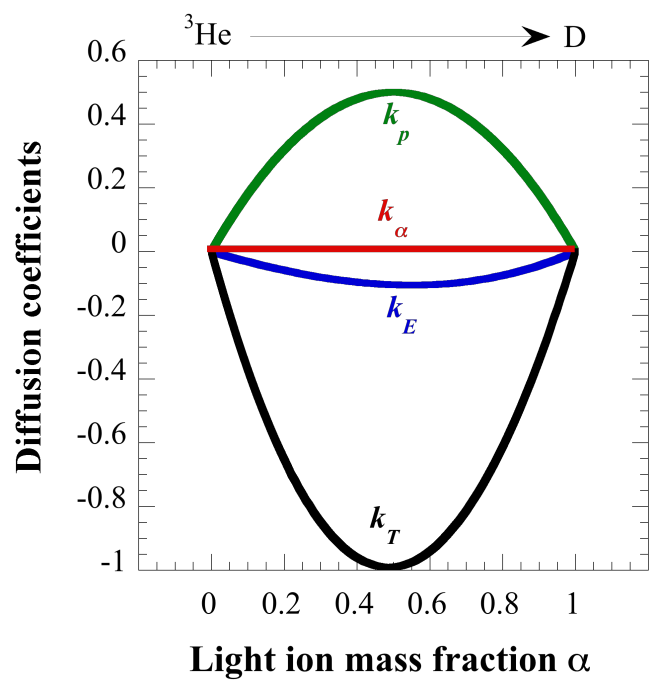

**Fig. 4**

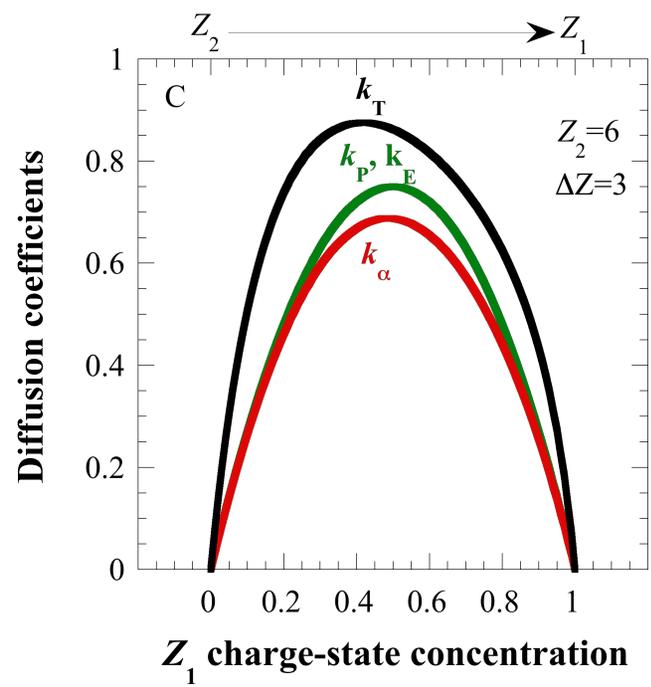